# Transfiguration of Quantum Walks on a line

Mahesh N. Jayakody and Asiri Nanayakkara*

*National Institute of Fundamental Studies,*

*Hanthana Road, Kandy, Sri Lanka*

## Abstract

We introduce an analytically treatable spin decoherence model for quantum walk on a line that yields the exact position probability distribution of an unbiased classical random walk at all-time scales. This spin decoherence model depicts a quantum channel in which simultaneous bit and phase flip operator is applied at random on the coin state. Based on this result we claim that there exist certain quantum channels that can produce exact classical statistical properties for a given one-dimensional quantum walk. Moreover, from the perspective of quantum computing, decoherence model introduced in this study may have useful algorithmic applications when it is applied on quantum walks with non-local initial states.

1. Introduction

Quantum walks (QWs) which differ much from the classical counterparts have gained a profound attention of the scientific community by becoming an effective testing ground in various areas of science. Primarily, QWs contribute to theoretical and practical advancement in quantum algorithms [1-4] and quantum computing [5, 6]. In addition, they are used to model the transport in biological systems [7-9] and other physical phenomena such as Anderson localization [10-14] and topological phases [15, 16].

Quantum to classical transition in finite dimensional quantum walks has been a topic of great interest for some time [17-34]. In general, quantum to classical transition is triggered by launching interventions on the internal degree of freedom (coin state) [17, 21-27, 31, 34 and 36] or on the external degree of freedom (position state) [18, 29, 30, 33 and 34] or on both degrees of freedom [19, 20, 29 and 34] of the coin-walker system. Most often, such interventions are introduced in the form of a decoherence model. In this study we are particularly interested in decoherence models defined on coin degree of freedom that trigger quantum to classical transition in quantum walks on a line. The key signature that distinguish a classical walk from a quantum walk is the second central moment or the variance ($\sigma^2 = \langle x^2 \rangle_t - \langle x \rangle_t^2$) of the position probability distribution of the walk. Variance of a classical random walk grows linearly with time ($\sigma^2 \propto t$) [35]. In contrast, variance of a quantum walk grows quadratically with time ($\sigma^2 \propto t^2$) [28, 32]. Hence, as a rule of thumb, time dependency of the variance is utilized+ as a quantitative measure of quantum to classical transition in quantum walks.

In literature, it can be recognized some possible routes that yield classical signature in quantum walks as a consequence of interventions on the coin degree of freedom. The very first practice of such an effort can be found in [17] where a random element is introduced into the coin transformation to switch the walker from quantum to classical regime. Most obvious way of achieving the classical probabilities in site occupation of the quantum walker is, performing a strong projective measurement on the coin state of the walker at each step. The record of measurement outcomes always corresponds to a particular classical path. By averaging over all possible measurement outcomes, one can observe the emergence of classical signature in quantum walks [23]. Sometimes this is termed as the corresponding classical walk of a given quantum walk. Classical signature in quantum walks can be recovered even without performing a measurement on the coin state at each time step. Instead, one can replace the quantum coin with a new coin operator at every time step. As a result of using different coins at each step, the effect of interference between paths tends to diminish with time. After $n$ steps we have an accumulation of $n$ coins which are entangled with the positon of the walker. By performing a measurement on the coin state of the final coin-walker state, it is possible to acquire a unique classical path. The expected classical result can be produced by averaging over all such possible outcomes [22]. In some studies, [20, 29] various choices of projections on coin space are used to break the coherence in the system partially and give rise to classical-like behavior. Reference [27] shows that quantum walks changeover between quantum and classical regime depending on the strength of the measurement made on the coin state of the walker. The measurement strength is determined via an ancilla that serves as a coin meter. The studies in [26,31] have achieved quantum to classical transition by introducing random fluctuations in the coin degree of freedom. A microscopic decoherence model for Hadamard walk is introduced in [24] by attaching a stochastic Hermitian operator to the single-qubit chirality space. Under this decoherence model, drifts and fluctuations occur in the parameters of the total Hamiltonian and as a consequence, Hadamard walk is exposed to a unitary noise and gives classical-like behavior in the long time limit. Phase-space approach in quantum walks has been used to study the behavior of a quantum walker when the quantum coin interacts with the environment [25]. In

[38] the phase-space approach is used to analyze the quantum walk scheme of a single cesium atom moving along a one dimensional optical lattice under the influence of a spin decoherence model. Space and time depended coin operators are used in [36, 37] to achieve classical-like behavior in quantum walks. A phenomenological decoherence model is introduced in [21, 23] for quantum walks on a line by defining a complete positive map on the coin degree of freedom. This model contains decoherence schemes like pure dephasing and weak measurements on coin space and characterizes a quantum walk with a coin subjected to decoherence. Analytical solution for a special case of this model has been derived for the usual Hadamard walk and hence it has been proved that quantum to classical transition of the Hadamard walk occurs only in the asymptotic limit. In [34] the aforementioned phenomenological decoherence model is generalized to all kinds of decoherence that includes coin, position and coin-position decoherence.

We consider a special case of the general decoherence model given in [21, 23]. The significance of the model discussed in this paper is that it is analytically treatable for any one-dimensional quantum walk scheme. Therefore, instead of deriving formulas for dispersion of the position probability distribution, we show that the model we consider here can produce the probability density function of a classical unbiased random walk at all-time scales. It can be identified two key features common to most decoherence models defined on the coin degree of freedom. One is, when the time step is fixed, the perfect quantum to classical transition can be achieved only by increasing the rate of decoherence [20, 21, 23, 24, and 29]. On the other hand, when there is no constrains on the time, even for low decoherence rates, perfect quantum to classical transition can be achieved in the asymptotic limit [21, 23, and 24]. In contrast to these behaviors, the decoherence model studied in this paper can give exact classical distribution under a fixed decoherence rate ($p = 1/2$) at all-time scales. Furthermore, we state that model introduced in this study may have useful algorithmic applications when it is applied on quantum walk with a non-local initial state.

The paper is organized in the following way. In section 2 a brief introduction on quantum walk is given along with its mathematical framework. Section 3 and 4 are devoted to formulate the spin decoherence model. In section 5, we discuss about one possible application of this spin decoherence model.

## 2. Quantum walk on a line

Let us consider the standard model of Quantum walk on line which comprises a two-state coin and a walker. Evolution of the coin-walker system is governed by a unitary operator $U$ defined on the tensor product of two Hilbert spaces, $H_x \otimes H_c$ which are spanned by the position basis $\{|x\rangle\}_{x \in Z}$ and the coin basis $\{|c\rangle\}_{c \in \{R,L\}}$. Single-step progression of the system is a sequential process in which the coin is tossed at first and then the walker is moved either to left ($L$) or right ($R$) conditional upon the outcome of the coin. In the language of mathematics, progression of the system can be represented by a combination of a shift operator ($S$) that acts on the position of the particle, two projectors ($P_R$ and $P_L$) on coin Hilbert Space that forms a complete orthogonal system of complimentary projections and a coin operator ($C$). Hence, single-step evolution operator of the quantum walk is given by

$$U = (S \otimes P_R + S^\dagger \otimes P_L) \cdot (\mathbb{I} \otimes C) \quad (1)$$

where $S$ and $S^\dagger$ are unitary shift operators defined by

$$S|x\rangle = |x+1\rangle \qquad S^\dagger|x\rangle = |x-1\rangle \quad (2)$$

and $P_R$ and $P_L$ are orthogonal projectors such that $P_R + P_L = \mathbb{I}$. For the sake of simplicity let us conduct our analysis in momentum space ($H_k$). Transformation from position basis to $k$-basis is given by

$$|k\rangle = \sum_x e^{ikx} |x\rangle \quad (3)$$

where $k \in [-\pi, \pi)$. The inverse transformation is given by

$$|x\rangle = \int_{-\pi}^{\pi} \frac{dk}{2\pi} e^{-ikx} |k\rangle \quad (4)$$

Observe that, $S|k\rangle = \sum_x e^{ikx} S|x\rangle = \sum_x e^{ikx} |x+1\rangle = e^{-ik} \sum_x e^{ikx} |x\rangle = e^{-ik} |k\rangle$. In the same fashion, $S^\dagger |k\rangle = e^{ik} |k\rangle$. Now let us rewrite the evolution operator $U$ in $k$-basis. Let $|k\rangle \otimes |\phi\rangle$ be an arbitrary state where $|k\rangle \in H_k$ and $|\phi\rangle \in H_c$. Then,

$$U(|k\rangle \otimes |\phi\rangle) = (\mathbb{I} \otimes (e^{-ik} P_R + e^{ik} P_L) C)(|k\rangle \otimes |\phi\rangle) \quad (5)$$

Thus in momentum space, $U$ is represented by the unitary operator

$$U_k = C_k \tag{6}$$

where $C_k = (e^{-ik}P_R + e^{ik}P_L)C$ and $P_L + P_R = \mathbb{I}$. This representation allows us to view a single-step evolution of an arbitrary state of the coin-walker system as a transformation performed only on the coin state of that arbitrary state.

3. Quantum walk under decoherence

Consider a one dimensional quantum walk scheme in which a Complete Positive Map (CPM) is performed on the coin degree of freedom at each step before performing the unitary flip of the coin. CPM is defined in such a way that it introduces a probability mixture of unitary transformations on the coin degree of freedom. Such a scheme resembles a quantum walk with a single coin subjected to decoherence. Probability mixture in CPM directly gives rise to a mixed quantum state and hence urges the need of employing the density operators of quantum sates in the analysis.

Let $\{\hat{A}_n\}_n$ be a set of operators on coin degree of freedom that satisfy the condition $\sum_n \hat{A}_n^\dagger \hat{A}_n = \mathbb{I}$. We can write the general density operator of the coin-walker system in the $k$-basis that mimics the aforementioned scheme by following the analysis given in [21,23]. Suppose the walker commences the walk at origin. Then the density operator of the initial state can be written as

$$\rho(t=0) = \int_{-\pi}^{\pi} \frac{dk}{2\pi} \int_{-\pi}^{\pi} \frac{dk'}{2\pi} |k\rangle\langle k'| \otimes |\phi_0\rangle\langle\phi_0| \tag{7}$$

where $|\phi_0\rangle\langle\phi_0|$ is the density operator for the initial coin state. After a single step the state becomes;

$$\rho(t=1) = \int_{-\pi}^{\pi} \frac{dk}{2\pi} \int_{-\pi}^{\pi} \frac{dk'}{2\pi} |k\rangle\langle k'| \otimes \sum_n C_k \hat{A}_n |\phi_0\rangle\langle\phi_0| \hat{A}_n^\dagger C_{k'}^\dagger \tag{8}$$

The state at time $t$ can be written as,

$$\rho(t) = \int_{-\pi}^{\pi} \frac{dk}{2\pi} \int_{-\pi}^{\pi} \frac{dk'}{2\pi} |k\rangle\langle k'| \otimes \sum_{n_1,\ldots,n_t} C_k \hat{A}_{n_t} \ldots C_k \hat{A}_{n_1} |\phi_0\rangle\langle\phi_0| \hat{A}_{n_1}^\dagger C_{k'}^\dagger \ldots \hat{A}_{n_t}^\dagger C_{k'}^\dagger \quad (9)$$

In terms of a superoperator $\mathcal{L}_{k,k'}$ defined on the coin degree of freedom we can write the state of the coin-walker system at time $t$ in a compact form as follows

$$\rho(t) = \int_{-\pi}^{\pi} \frac{dk}{2\pi} \int_{-\pi}^{\pi} \frac{dk'}{2\pi} |k\rangle\langle k'| \otimes \mathcal{L}_{k,k'}^t |\phi_0\rangle\langle\phi_0| \quad (10)$$

4. Emergence of Classical Behavior

In this section we introduce a special type of CPM defined on the coin degree of freedom. This complete positive map has the ability to generate exact classical signature at all-time scales. We present it in the form of a proposition and provide the proof along with it.

**Proposition:**
*For each scheme of one dimensional Quantum walks that commences from a single position, a quantum channel can be defined in such a way that the position distribution of the quantum walk is equivalent to that of a classical unbiased random walk at all-time scales.*

For the sake of simplicity we take the initial position as $x_0 = 0$. If the statement holds for $x_0 = 0$ then it can be generalized into any position simply by displacing the coordinate system.

Let us consider the standard quantum walk on a line. General form of the coin operator [28] that governs the walk can be written as

$$C = \begin{pmatrix} \cos(\theta) & e^{i\phi_1}\sin(\theta) \\ e^{i\phi_2}\sin(\theta) & -e^{i(\phi_1+\phi_2)}\cos(\theta) \end{pmatrix} \quad (11)$$

where $\theta \in [0, 2\pi)$, and $\phi_1, \phi_2 \in [0, \pi)$.

Now let us define two operators on coin Hilbert space as

$$A_0 = \sqrt{p}\left(e^{i\phi_3}|R\rangle\langle L| - e^{-i\phi_3}|L\rangle\langle R|\right)$$
$$A_1 = \sqrt{1-p}\,\mathbb{I} \tag{12}$$

where $0 < p < 1$ and $\phi_3 \in [0, \pi)$. Note that $\sum_n \hat{A}_n^\dagger \hat{A}_n = \mathbb{I}$. We can introduce decoherence into coin degree of freedom by defining a complete positive map using $A_0$ and $A_1$. In this context $p$ represent the probability of a decoherence event happening per time step. Write $|R\rangle = (1\ \ 0)^T$ and $|L\rangle = (0\ \ 1)^T$. Then $P_R = |R\rangle\langle R|$ and $P_L = |L\rangle\langle L|$ become two orthogonal projectors in coin Hilbert space such that $P_L + P_R = \mathbb{I}$. From (6) we can define $C_k$ as

$$C_k = \begin{pmatrix} e^{-ik}\cos(\theta) & e^{-i(k-\phi_1)}\sin(\theta) \\ e^{i(k+\phi_2)}\sin(\theta) & -e^{i(k+\phi_1+\phi_2)}\cos(\theta) \end{pmatrix} \tag{13}$$

Let $O$ be the density operator for the initial coin state. From (8) the action of the superoperator $\mathcal{L}_{k,k'}$ on $O$ can be written as

$$\mathcal{L}_{k,k'} O = \sum_{n=0}^{1} C_k \hat{A}_n O \hat{A}_n^\dagger C_{k'}^\dagger \tag{14}$$

A convenient representation of $O$ is given by

$$O = r_0\sigma_0 + r_1\sigma_1 + r_2\sigma_2 + r_3\sigma_3 \tag{15}$$

where $\sigma_0 = \mathbb{I}$ and $\sigma_{1,2,3} = \sigma_{x,y,z}$ are usual Paulin matrixes. This representation allows us to express $O$ as a column vector

$$O = (r_0\ \ r_1\ \ r_2\ \ r_3)^T \tag{16}$$

Then we can alternatively represent the superoperator $\mathcal{L}_{k,k'}$ which corresponds to the complete positive map containing $A_0$ and $A_1$ as;

$$\mathcal{L}_{k,k'}O = \begin{pmatrix} a_{11} & a_{12} & a_{13} & a_{14} \\ 0 & a_{22} & a_{23} & a_{24} \\ 0 & a_{32} & a_{33} & a_{34} \\ a_{41} & a_{42} & a_{43} & a_{44} \end{pmatrix} \begin{pmatrix} r_0 \\ r_1 \\ r_2 \\ r_3 \end{pmatrix} \tag{17}$$

$a_{11} = \cos(k - k')$

$a_{12} = i\,[(p-1)\cos(\phi_1) + p\cos(\phi_1 - 2\phi_3)]\sin(k-k')\sin(2\theta)$

$a_{13} = i\sin(k-k')\sin(2\theta)\,[p\sin(\phi_1 - 2\phi_3) - (p-1)\sin(\phi_1)]$

$a_{14} = i\,(2p-1)\cos(2\theta)\sin(k-k')$

$a_{22} = \cos(2\theta)\cos(k+k'+\phi_2)[(p-1)\cos(\phi_1) + p\cos(\phi_1 - 2\phi_3)]$
$\quad\quad -\sin(k+k'+\phi_2)[(p-1)\sin(\phi_1) + p\sin(\phi_1 - 2\phi_3)]$

$a_{23} = [p\cos(\phi_1 - 2\phi_3) - (p-1)\cos(\phi_1)]\sin(k+k'+\phi_2)$
$\quad\quad +\cos(2\theta)\cos(k+k'+\phi_2)\,[p\sin(\phi_1 - 2\phi_3) - (p-1)\sin(\phi_1)]$

$a_{24} = -(2p-1)\cos(k+k'+\phi_2)\sin(2\theta)$

$a_{32} = \cos(2\theta)[(p-1)\cos(\phi_1) + p\cos(\phi_1 - 2\phi_3)]\sin(k+k'+\phi_2)$
$\quad\quad +\cos(k+k'+\phi_2)\,[(p-1)\sin(\phi_1) + p\sin(\phi_1 - 2\phi_3)]$

$a_{33} = -\cos(k+k'+\phi_2)[p\cos(\phi_1 - 2\phi_3) - (p-1)\cos(\phi_1)]$
$\quad\quad +\cos(2\theta)\sin(k+k'+\phi_2)\,[p\sin(\phi_1 - 2\phi_3) - (p-1)\sin(\phi_1)]$

$a_{34} = -(2p-1)\sin(2\theta)\sin(k+k'+\phi_2)$

$a_{41} = -i\sin(k-k')$

$a_{42} = -\cos(k-k')[(p-1)\cos(\phi_1) + p\cos(\phi_1 - 2\phi_3)]\sin(2\theta)$

$a_{43} = -\cos(k-k')\sin(2\theta)\,[p\sin(\phi_1 - 2\phi_3) - (p-1)\sin(\phi_1)]$

$a_{44} = -(2p-1)\cos(k-k')\cos(2\theta)$

Suppose $p = 1/2$. This resembles a CPM that provides an equal chance for the coin degree of freedom to experience decoherence either by $A_0$ or $A_1$ operator. Furthermore, observe that by choosing $\phi_3 = \phi_1$ we can simplify $\mathcal{L}_{k,k'}$ into a more convenient form. Then the corresponding superoperator can be written as;

$$\mathcal{L}_{k,k'}O = \begin{pmatrix} a_{11} & 0 & 0 & 0 \\ 0 & a_{22} & a_{23} & 0 \\ 0 & a_{32} & a_{33} & 0 \\ a_{41} & 0 & 0 & 0 \end{pmatrix} \begin{pmatrix} r_0 \\ r_1 \\ r_2 \\ r_3 \end{pmatrix} \tag{18}$$

where

$a_{11} = cos(k - k')$

$a_{22} = sin(k + k' + \phi_2)sin(\phi_1)$

$a_{23} = sin(k + k' + \phi_2)cos(\phi_1)$

$a_{32} = -cos(k + k' + \phi_2) sin(\phi_1)$

$a_{33} = -cos(k + k' + \phi_2)cos(\phi_1)$

$a_{41} = -i\, sin(k - k')$

Moreover, the time evolution of the operator $\mathcal{L}_{k,k'}$ is given by

$$\mathcal{L}_{k,k'}^t O = \begin{pmatrix} a_{11}(t) & 0 & 0 & 0 \\ 0 & a_{22}(t) & a_{23}(t) & 0 \\ 0 & a_{32}(t) & a_{33}(t) & 0 \\ a_{41}(t) & 0 & 0 & 0 \end{pmatrix} \begin{pmatrix} r_0 \\ r_1 \\ r_2 \\ r_3 \end{pmatrix} \qquad (19)$$

where $t \in \mathbb{N}$ and

$a_{11}(t) = cos^t(k - k')$

$a_{22}(t) = (-1)^{t+1} cos^{t-1}(k + k' + \phi_1 + \phi_2) sin(\phi_1) sin(k + k' + \phi_2)$

$a_{23}(t) = (-1)^{t+1} cos^{t-1}(k + k' + \phi_1 + \phi_2) cos(\phi_1) sin(k + k' + \phi_2)$

$a_{32}(t) = (-1)^t cos^{t-1}(k + k' + \phi_1 + \phi_2) cos(k + k' + \phi_2) sin(\phi_1)$

$a_{33}(t) = (-1)^t cos^{t-1}(k + k' + \phi_1 + \phi_2) cos(\phi_1) cos(k + k' + \phi_2)$

$a_{41}(t) = -i\, cos^{t-1}(k - k') sin(k - k')$

By rewriting $\mathcal{L}_{k,k'}^t O$ in Pauli matrix basis we get

$$\mathcal{L}_{k,k'}^t O = \begin{pmatrix} c_{11}(t) & c_{12}(t) \\ c_{21}(t) & c_{22}(t) \end{pmatrix} \qquad (20)$$

where

$c_{11}(t) = r_0 e^{-i(k-k')} cos^{t-1}(k - k')$

$c_{12}(t) = -ie^{-i(k+k'+\phi_2-\pi t)} cos^{t-1}(k + k' + \phi_1 + \phi_2)[r_1 sin(\phi_1) + r_2 cos(\phi_1)]$

$c_{21}(t) = ie^{i(k+k'+\phi_2+\pi t)} cos^{t-1}(k + k' + \phi_1 + \phi_2)[r_1 sin(\phi_1) + r_2 cos(\phi_1)]$

$c_{22}(t) = r_0 e^{i(k-k')} cos^{t-1}(k - k')$

Now let us trace out the coin space from the density matrix of the coin-walker system given in (10). Then we can write the position density matrix as follows;

$$\rho_w(t) = Tr_c[\rho(t)] = \int_{-\pi}^{\pi} \frac{dk}{2\pi} \int_{-\pi}^{\pi} \frac{dk'}{2\pi} |k\rangle\langle k'| \times Tr(\mathcal{L}_{k,k'}^t |\phi_0\rangle\langle\phi_0|) \qquad (21)$$

Trace of $\mathcal{L}_{k,k'}^t |\phi_0\rangle\langle\phi_0|$ can be determined from (20). Hence we can rewrite $\rho_w(t)$ in position basis as;

$$\rho_w(t) = \frac{2r_0}{(2\pi)^2} \sum_{x,x'} \int_{-\pi}^{\pi} dk \int_{-\pi}^{\pi} dk' \, e^{ikx} e^{-ik'x'} \cos^t(k - k') |x\rangle\langle x'| \qquad (22)$$

Definite integrals given in (22) can be evaluated by expressing the cosine function in terms of complex exponential function and then by applying the binomial theorem. In doing so we get an expression for $\rho_w(t)$ as;

$$\rho_w(t) = \frac{2r_0}{2^t} \sum_{x,x'} \sum_{s=0}^{t} {}_s^tC \left(\frac{\sin[\pi(x + t - 2s)]}{\pi(x + t - 2s)}\right)\left(\frac{\sin[\pi(x' + t - 2s)]}{\pi(x' + t - 2s)}\right) |x\rangle\langle x'| \qquad (23)$$

where ${}_s^tC = \frac{t!}{s!(t-s)!}$

From the properties of sine function, it is possible to argue that all the terms in (23) except $x = x' = 2s - t$ term become zero. In addition, we have $2r_0 = Tr(0) = 1$. Thus the explicit form of $\rho_w(t)$ can be written as;

$$\rho_w(t) = \sum_x \frac{t!}{2^t \left(\frac{t+x}{2}\right)! \left(\frac{t-x}{2}\right)!} |x\rangle\langle x| \qquad (24)$$

where $x \in \{-t, -t+2, \ldots, t-2, t\}$

According to the expression given in (24) the off-diagonal terms of the position density matrix have vanished. In other words, the terms that contained information of coherence among the position states have become zero. This represents a quantum system subjected to perfect decoherence. In addition, note that $\rho_w(t)$ has no dependence on the initial coin state or the coin operator of the quantum walk. Now let us determine the probability of finding the walker at point $y$ at time $t$ in the following way

$$p(y,t) = \langle y|\rho_w(t)|y\rangle = \frac{t!}{2^t \left(\frac{t+y}{2}\right)! \left(\frac{t-y}{2}\right)!} \tag{25}$$

Note that the expression in (25) exactly equals to the probability mass function of an unbiased classical random walk [24, 36]. Thus this completes the proof.

One would be interested in the experimental realization of this spin decoherence scheme. Consider the Kraus operator $A_0$ given in (12). Assume $\phi_3 = \phi_1$ From the matrix product of the general coin operator C given in (11) and $A_0$ we have a new coin operator of the form

$$D = \begin{pmatrix} -\sin(\theta) & e^{i\phi_1}\cos(\theta) \\ e^{i\phi_2}\cos(\theta) & e^{i(\phi_1+\phi_2)}\sin(\theta) \end{pmatrix} \tag{26}$$

By altering the coin operators $C$ and $D$ at random one can generate a quantum walk on line which ultimately gives a position probability distribution of an unbiased classical random walk. We present a numerical simulation for the standard Hadamard coin walk under this spin decoherence model for further justification. Choose $\phi_1 = \phi_2 = \phi_3 = 0$, $\theta = \pi/4$ and $p = 1/2$. From (12) we have

$$\tilde{A}_0 = \frac{1}{\sqrt{2}}\begin{pmatrix} 0 & 1 \\ -1 & 0 \end{pmatrix} \qquad \tilde{A}_1 = \frac{1}{\sqrt{2}}\begin{pmatrix} 1 & 0 \\ 0 & 1 \end{pmatrix} \tag{27}$$

Let the initial state be $|\psi_{in}\rangle = |0\rangle_c \otimes |0\rangle_x$. In performing the simulation we apply operator $\sqrt{2}\tilde{A}_0$ or $\sqrt{2}\tilde{A}_1$ in (27) with equal probability on the coin state. Note that $\sqrt{2}\tilde{A}_0$ and $\sqrt{2}\tilde{A}_1$ are unitary operators. Thus, there is no need of renormalizing the state. Afterwards, the mean probability distribution was calculated by averaging over 1000 trials. Figure I (a) and (b) show the final probability distribution of classical random walk and quantum walk with decoherence for 100 time steps respectively. Theoretically predicated position probability distribution (dotted line) is also plotted in the same figure for comparison.

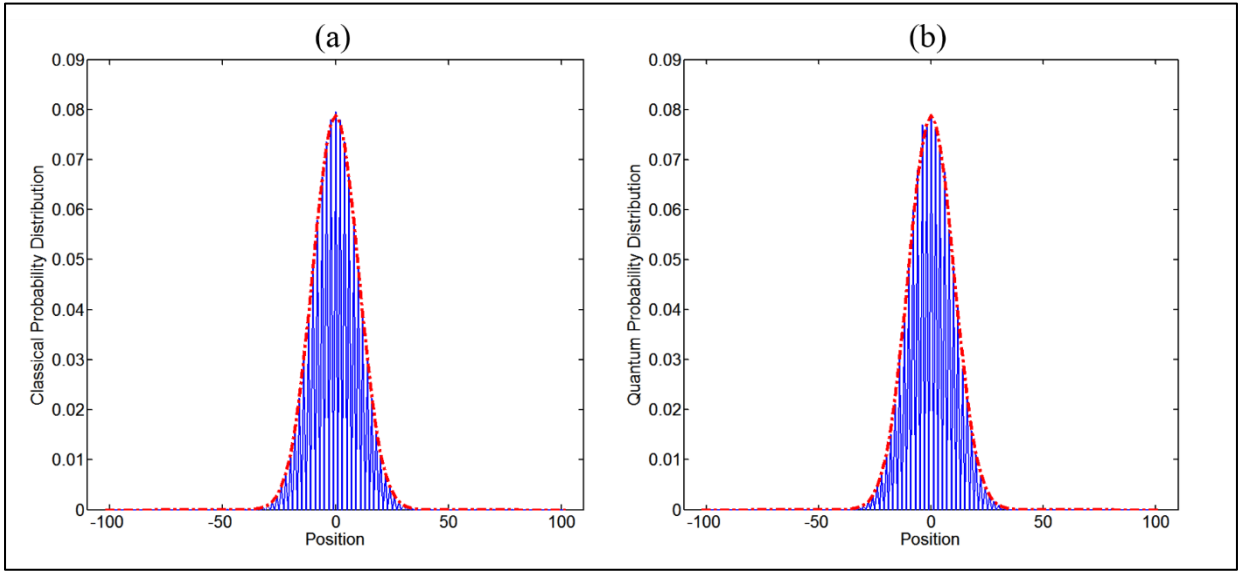

Figure I: Position probability distributions for 100 times steps (a) Classical Random Walk ($\sigma_{CRW}$=10) (b) Quantum walk for the initial state $|\psi_{in}\rangle = |0\rangle_x \otimes |0\rangle_c$ and $\phi_1 = \phi_2 = 0$, $\theta = \pi/4$, $p = 1/2$ ($\sigma_{QW}$=9.9732) and Dotted line: Theoretical distribution

5. QW with a non-local initial state under decoherence

It is interesting to test the spin decoherence model given in (12) on a quantum walk that starts with a non-local initial state. This scenario can be viewed in another way. Suppose a one dimensional quantum walk is allowed to evolve stating from an arbitrary state without any decoherence effects. After certain time steps the spin decoherence model given in (12) is introduced to the system. Let us determine the position probability distribution of the walk under this situation.

Let the initial state of the coin-walker system be

$$|\psi_0\rangle = \sum_x [|x\rangle \otimes (a_x|R\rangle + b_x|L\rangle)] \tag{28}$$

where $\sum_x |a_x|^2 + |b_x|^2 = 1$

The density matrix that represents the initial coin-walker system is given by

$$\rho_0 = |\psi_0\rangle\langle\psi_0| = \sum_{x,x'}|x\rangle\langle x'| \otimes \left(a_x a_{x'}^*|R\rangle\langle R| + a_x b_{x'}^*|R\rangle\langle L| + b_x a_{x'}^*|L\rangle\langle R| + b_x b_{x'}^*|L\rangle\langle L|\right) \quad (29)$$

In momentum representation $\rho_0$ takes the following from

$$\rho_0 = \sum_{x,x'}\int_{-\pi}^{\pi}\frac{dk}{2\pi}\int_{-\pi}^{\pi}\frac{dk'}{2\pi}e^{-ikx}e^{ik'x'}|k\rangle\langle k'| \otimes \left(a_x a_{x'}^*|R\rangle\langle R| + a_x b_{x'}^*|R\rangle\langle L| + b_x a_{x'}^*|L\rangle\langle R| + b_x b_{x'}^*|L\rangle\langle L|\right) \quad (30)$$

From (10) the state after time $t$ can be written as

$$\rho_t = \sum_{x,x'}\int_{-\pi}^{\pi}\frac{dk}{2\pi}\int_{-\pi}^{\pi}\frac{dk'}{2\pi}e^{-ikx}e^{ik'x'}|k\rangle\langle k'| \otimes \mathcal{L}_{k,k'}^t\left(a_x a_{x'}^*|R\rangle\langle R| + a_x b_{x'}^*|R\rangle\langle L| + b_x a_{x'}^*|L\rangle\langle R| + b_x b_{x'}^*|L\rangle\langle L|\right) \quad (31)$$

Since $\mathcal{L}_{k,k'}$ is linear, operation of addition is preserved under $\mathcal{L}_{k,k'}$ and hence we can apply the super operator on each outer product of the coin space separately. Then by tracing out the coin space from the density matrix of the coin-walker system given in (31) we can write the position density matrix as follows;

$$\rho_w(t) = Tr_c(\rho_t) = \sum_{x,x'}\int_{-\pi}^{\pi}\frac{dk}{2\pi}\int_{-\pi}^{\pi}\frac{dk'}{2\pi}e^{-ikx}e^{ik'x'}|k\rangle\langle k'|$$

$$\times \left(a_x a_{x'}^* Tr(\mathcal{L}_{k,k'}^t|R\rangle\langle R|) + a_x b_{x'}^* Tr(\mathcal{L}_{k,k'}^t|R\rangle\langle L|) + b_x a_{x'}^* Tr(\mathcal{L}_{k,k'}^t|L\rangle\langle R|) + b_x b_{x'}^* Tr(\mathcal{L}_{k,k'}^t|L\rangle\langle L|)\right) \quad (32)$$

Note that from (20), for each initial coin state $O$, the trace of the operator $\mathcal{L}_{k,k'}^t O$ can be written as $Tr(\mathcal{L}_{k,k'}^t O) = 2r_0 \cos^t(k - k')$ where $r_0$ is a coefficient corresponding to the coin state $O$ under the Paulin basis expansion. Using this result it can be easily shown that $Tr(\mathcal{L}_{k,k'}^t|R\rangle\langle L|) = Tr(\mathcal{L}_{k,k'}^t|L\rangle\langle R|) = 0$ and $Tr(\mathcal{L}_{k,k'}^t|R\rangle\langle R|) = Tr(\mathcal{L}_{k,k'}^t|L\rangle\langle L|) = \cos^t(k - k')$. Thus (32) can be rewritten as

$$\rho_w(t) = \sum_{x,x'}\int_{-\pi}^{\pi}\frac{dk}{2\pi}\int_{-\pi}^{\pi}\frac{dk'}{2\pi}e^{-ikx}e^{ik'x'}(a_x a_{x'}^* + b_x b_{x'}^*)\cos^t(k - k')|k\rangle\langle k'| \quad (33)$$

The probability of finding the walker at point $y$ at time $t$ is given by

$$p(y,t) = \langle y|\rho_w(t)|y\rangle = \sum_{x,x'} \int_{-\pi}^{\pi} \frac{dk}{2\pi} \int_{-\pi}^{\pi} \frac{dk'}{2\pi} (a_x a_{x'}^* + b_x b_{x'}^*) e^{-ik(x-y)} e^{ik'(x'-y)} \cos^t(k-k') \quad (34)$$

Expression given in (34) can be evaluated by expressing the cosine function in terms of complex exponential function and then by applying the binomial theorem. In doing so we get an expression for $p(y,t)$ as

$$p(y,t) = \sum_x (|a_x|^2 + |b_x|^2) \frac{t!}{2^t \left(\frac{t+x+y}{2}\right)! \left(\frac{t-x-y}{2}\right)!} \quad (35)$$

The result obtained in (35) can be interpreted as a convex superposition of $x$ number of independent unbiased classical random walks. One can look at this result in another perspective. Consider an unbiased classical random walk in which the initial position $x$ of the walker is selected according to probabilities $p_x = |a_x|^2 + |b_x|^2$. After choosing the initial position, the walker is allowed to move in the usual unbiased classical random walk. The position probability distribution of such walk is exactly the same as the result given in (35). Hence, by adopting the decoherence model introduced in this study, one can develop a quantum walk to model the aforementioned classical scenario.

6. Conclusion

We introduce a specific type of complete positive map defined on coin degree of freedom. This CPM produces decoherence in the coin degree of freedom and results in transition from quantum to classical regime for any scheme of quantum walk on the line. The significance of this transition is that it produces exact classical unbiased random walk. Most decoherence models introduce interventions on diagonal terms of the density operator and trigger the quantum to classical transition eventually. On the contrary, the decoherence model discussed in here introduces interventions on off-diagonal terms and produce the exact classical result at all-time scales.